\documentstyle[graphicx,longtable,amsmath]{aipproc}

\newcommand{\bea}{\begin{eqnarray}}
\newcommand{\eea}{\end{eqnarray}}

\newcommand{\pa}{\partial}
\newcommand{\cm}{{\rm cm}}

\newcommand{\const}{{\rm const}}
\newcommand{\rms}{{\rm rms}}
\newcommand{\localepsfilespath}{eps_files}
\renewcommand{\Re}{{\rm Re}}
\renewcommand{\Im}{{\rm Im}}

\renewcommand{\vec}[1]{\textnormal{\boldmath$#1$}}

\begin{document}
\title{Wake and Impedance}

\author{G. V. Stupakov}
\address{Stanford Linear Accelerator Center
Stanford University, Stanford, CA 94309}

\maketitle

\section{Introduction}
In this lecture we will develop a concept of wakes and impedances for
relativistic beams interacting with the surrounding environment.
Among the numerous publications and reviews on this subject, we refer
here to recent books \cite{chao93,zotter98k,chao99t}, where the
reader can find a more detailed treatment and further references.

We will use the CGS system of units throughout this paper.

\section{Interaction of Moving Charges in Free Space}\label{vacuum}

We begin with interactions of particles that moving with constant
velocity in free space. If the material walls are far from the
particles, their effect in the first approximation can be neglected.

\begin{figure}[ht]
        \begin{center}
        \includegraphics[scale=0.7]{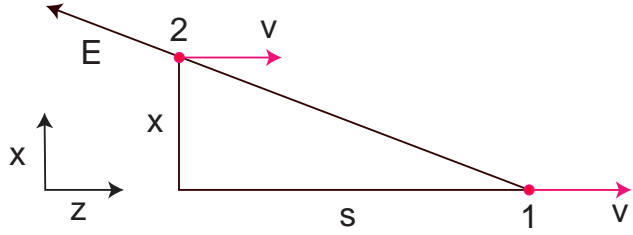}
        \end{center}
        \caption{A leading particle 1 and a trailing particle
        2 traveling in free space with parallel velocities $\vec{v}$. Shown
        also is the coordinate system $x,z$.
        \label{moving_charges}
        }
        \end{figure}

Let us consider a \emph{leading} particle of charge $q$ moving with
velocity $v$, and a \emph{trailing} particle of \emph{unit} charge
moving behind the leading one on a parallel path at a distance $s$
with an offset $x$, as shown in Fig. \ref{moving_charges}. We want to
find the force which the leading particle exerts on the trailing one.

We will use the following expressions for the electric and magnetic
fields of a particle moving with a constant velocity (see,
e.g.,\cite{landau_lifshitz_ctf}):
    \begin{equation} \label{field_in_vacuum}
    \vec{E}=\frac{q\vec{R}}{\gamma^2 R^{*3}}\,,
    \qquad
    \vec{H}=\frac{1}{c}\vec{v}\times\vec{E}\,,
    \end{equation}
where $\vec{R}$ is the vector drawn 
from point 1 to point 2, $R^{*2}=s^2+x^2/\gamma^2$, and $\gamma =
\left(1-v^2/c^2\right)^{-1/2}$.

From Eq. (\ref{field_in_vacuum}) we find that the longitudinal force
acting on the trailing charge is
    \begin{equation}
    F_l=E_z = -\frac{qs}{\gamma^2 (s^2+x^2/\gamma^2)^{3/2}}\,,
    \end{equation}
and the transverse force is
    \begin{equation}
    F_t=E_x-\frac{v}{c}B_y=\frac{qx}{\gamma^4 (s^2+x^2/\gamma^2)^{3/2}}\,.
    \end{equation}
In accelerator physics, the force $\vec{F}$ is often called \emph{the
space-charge force}.

It is easy to see that for any position given by $s$ and $x$, the
longitudinal force decreases with the growth of $\gamma$ as
$\gamma^{-2}$. For the transverse force, if $s\gg x/\gamma$, $F_t\sim
\gamma^{-4}$, but for $s=0$, $F_t\sim \gamma^{-1}$. Hence, in the
limit of ultrarelativistic particles moving parallel to each other,
$\gamma \rightarrow \infty$, the electromagnetic interaction in free
space vanishes.

In this lecture, we will focus on the case of ultrarelativistic
charges, where $v \rightarrow c$. The space-charge effects discussed
above disappear in this limit, and the interaction between the
particles is  due only to the presence of material walls.

Note that, taking the limit $v \rightarrow c$ in
Eq.~(\ref{field_in_vacuum}) and recalling that $s=vt-z$, we can write
the electromagnetic field of an ultrarelitivistic charge in free
space as
    \begin{equation} \label{field_in_vacuum_delta}
    \vec{E}=\frac{2q\vec{r}}{r^{2}}\delta(z-ct)
    \,,
    \qquad
    \vec{H}=\vec{\hat{z}}\times\vec{E}\,,
    \end{equation}
where $\vec{r}=\vec{\hat{x}}x+\vec{\hat{y}}y$ is a two-dimensional
radius vector in a cylindrical coordinate system ($\vec{\hat{x}}$ and
$\vec{\hat{y}}$ are the unit vectors in the directions of $x$ and
$y$, respectively).

\section{Particles Moving in a Perfectly Conducting Pipe}

\begin{figure}[ht]
        \begin{center}
        \includegraphics[scale=0.7]{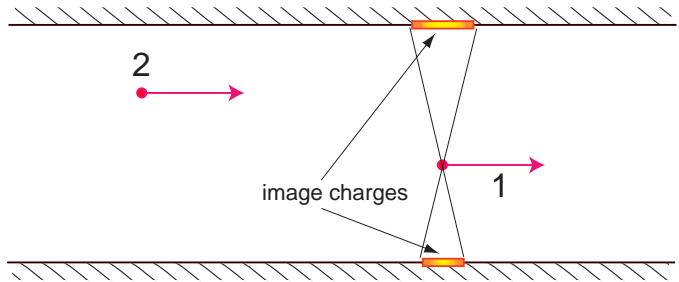}
        \end{center}
        \caption{Particles traveling inside a perfectly conducting
        pipe of arbitrary cross section. Shown are the image charges
        on the wall generated by the leading charge.
        \label{moving_charges_in_pipe}
        }
        \end{figure}

If particles from the above example move parallel to the axis in a
perfectly conducting cylindrical pipe of arbitrary cross section,
they induce image charges, on the surface of the wall, that screen
the metal from the electromagnetic field of the particles. The image
charges travel with the same velocity $v$ (see Fig.
\ref{moving_charges_in_pipe}). Since both the particles and the image
charges move on parallel paths, in the limit $v=c$, according to the
results in Section \ref{vacuum}, they do not interact with each other,
no matter how close to the wall the particles are.

Interaction between the particles in the ultrarelativistic limit can
occur if 1) the wall is not perfectly conducting, or 2) the pipe is
not cylindrical (which is usually due to the presence of RF cavities,
flanges, bellows, beam position monitors, slots, etc., in the vacuum
chamber).

%
%
%

\section{Causality and the ``Catch-Up'' Distance}

If a beam particle moves along a straight line with the speed of
light, the electromagnetic field of this particle scattered off the
boundary discontinuities will not overtake it and, furthermore, will
not affect the charges that travel ahead of it. The field can
interact only with the trailing charges in the beam that move behind
it. This constitutes the principle of \emph{causality} in the theory
of wakefields, according to which the interaction of a point charge
moving with the speed of light propagates only downstream and never
reaches the upstream part of the beam.

        \begin{figure}[ht]
        \begin{center}
        \includegraphics[scale=0.7]{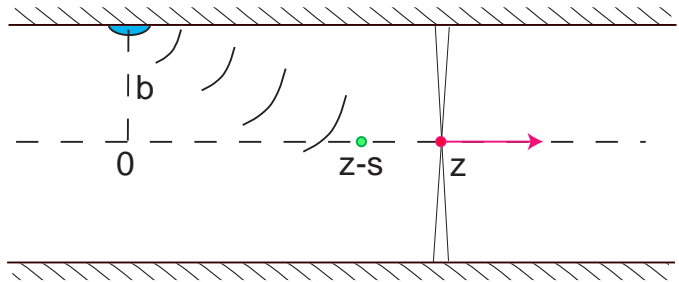}
        \end{center}
        \caption{A wall discontinuity located at $z=0$ scatters the
        electromagnetic field of an ultrarelativistic particle.
        When the particle moves to location $z$, the
        scattered field arrives to point $z-s$ .
        \label{catch_up}
        }
        \end{figure}
We can estimate the distance at which the electromagnetic field
produced by a leading charge reaches a trailing particles traveling
at a distance $s$ behind. Let us assume that a discontinuity located
at the surface of a pipe of radius $b$ at coordinate $z=0$ is passed
by the leading particle at time $t=0$, see Fig. \ref{catch_up}. If
the scattered field reaches point $s$ at time $t$, then
$ct=\sqrt{(z-s)^2+b^2}$, where $z$ is a coordinate of the leading
particle at time $t$, $z=ct$. Assuming that $s \ll b$, from these two
equations we find
    \begin{equation} \label{catch_up_distance}
    z\approx \frac{b^2}{2s}.
    \end{equation}
The distance $z$ given by this equation is often called the
\emph{catch-up distance}. Only after the leading charge has traveled
this distance away from the discontinuity, can a particle at point
$s$ behind it feel the wakefield generated by the discontinuity.

\section{Round Pipe with Resistive Walls}\label{res_wall_section}

Consider a round pipe of radius $b$, with finite wall conductivity
$\sigma$. A point charge moves along the $z$ axis of the pipe with
the speed of light, and a trailing particle follows the leading one
at a distance $s$. Both particles are assumed to be on the axis of
the pipe. Because of the symmetry of the problem, the only non-zero
component of the electromagnetic field on the axis is $E_z$, which,
depending on the sign, either accelerates or decelerates the trailing
charge. Our goal now is to find the field $E_z$ as a function of $s$.

If the conductivity of the pipe is large enough, we can use
perturbation theory to find the effect of the wall resistivity. In
the first approximation, we consider the pipe as a perfectly
conducting one. In this case, the electromagnetic field of the charge
is the same as in free space and is given by Eqs.
(\ref{field_in_vacuum_delta}). For what follows, we will need only
the magnetic field $H_\theta$,
    \begin{equation} \label{H_theta}
    H_\theta = \frac{2q}{r}\delta(z-ct).
    \end{equation}

Using the mathematical identity
    \begin{equation}
    \delta (z-ct)={1 \over 2\pi c} \int_{-\infty}^{\infty}d\omega
    e^{-i\omega(t-z/c)},
    \end{equation}
we will decompose $H_\theta$ into a Fourier integral,
    \begin{equation}
    H_\theta(r,z,t) =\int_{-\infty}^{\infty}d\omega
    H_{\theta\omega}(r)
    e^{-i\omega t+i\omega {z / c}},
    \end{equation}
where
    \begin{equation}\label{H_theta1}
    H_{\theta\omega}(r) = \frac{q}{\pi rc}.
    \end{equation}

In the limit where the skin depth $\delta$ corresponding to the
frequency $\omega$, $\delta = c/\sqrt{2\pi \sigma\omega}$, is much
smaller than the pipe radius , $\delta \ll b$, we can use the
Leontovich boundary condition \cite{landau_lifshitz_ecm} that
relates the tangential electric field $\vec{E}_t$ on the wall with the
magnetic one,
    \begin{equation} \label{Leontovich}
    \vec{E}_t=\zeta
    \vec{H}\times\vec{n},
    \end{equation}
where $\vec {n}$ is the unit vector normal to the surface and
directed toward the metal, and
    \begin{equation} \label{zeta}
    \zeta(\omega)
    =(1-i)\sqrt{\frac{\omega}{8\pi
    \sigma}}.
    \end{equation}
Combining Eqs.(\ref{Leontovich}), (\ref{zeta}) and (\ref{H_theta1}),
we find
    \begin{equation} \label{Ez}
    E_{z\omega}|_{r=b}
    =-(1-i)\sqrt{\frac{\omega}{8\pi
    \sigma}}
    \frac{q}{\pi b c}.
    \end{equation}

Equation (\ref{Ez}) gives us the longitudinal electric field on the
wall, but we need the field on the axis of the pipe. To find the
radial dependence of $E_{z\omega}$, we use Maxwell's equations, from
which it follows that the electric field in a vacuum satisfies the wave
equation. In the cylindrical coordinate system the wave equation for
$E_z$ is
    \bea
    && \frac{1}{c^2}
    \frac{\partial^2 E_z(r,z,t)}{\partial^2 t}
    - \Delta E_z(r,z,t)
    \nonumber \\
    &=&\frac{1}{c^2}
    \frac{\partial^2 E_z(r,z,t)}{\partial^2 t}
    - \frac{\partial^2 E_z(r,z,t)}{\partial^2 z}
    -
    \frac{1}{r}\frac{\partial }{\partial r}r\frac{\partial E_z(r,z,t)}{\partial r}
    = 0.
    \eea
Substituting the Fourier component
$E_{z\omega}(r)e^{-i\omega(t-z/c)}$ into this equation, we find
    \begin{equation}
    \frac{1}{r}\frac{\partial }{\partial r}r\frac{\partial E_{z\omega}}{\partial r}
    =0.
    \end{equation}
This equation has a general solution $E_{z\omega}=A+B\ln r$, where
$A$ and $B$ are arbitrary constants. Since we do not expect $E_z$ to
have a singularity on the axis, $B=0$. Hence the electric field does
not depend on $r$, $E_{z\omega} = \const$, and
    \begin{equation}
    E_{z\omega}|_{r=0}=E_{z\omega}|_{r=b},
    \end{equation}
implying that  $E_{z\omega}|_{r=0}$ is given by the same Eq.
(\ref{Ez}). Note that we have shown here that in the
ultrarelativistic case the longitudinal electric field inside the
pipe is constant throughout the pipe cross section.

To find $E_z(z,t)$ we make the inverse Fourier transformation,
    \begin{equation} \label{EzFourier}
    E_{z}(z,t)= \int_{-\infty}^{\infty}d\omega
    E_{z\omega}e^{-i\omega(t-z/c)},
    \end{equation}
which gives
    \begin{equation}\label{eq17}
    E_{z}(z,t)= (i-1)\frac{q}{\pi cb}
    \sqrt{\frac{1}{8\pi\sigma}}
    \int_{-\infty}^{\infty}d\omega
   \sqrt{{\omega}}e^{-i\omega(t-z/c)}.
    \end{equation}
The last integral can be taken analytically in the complex plane (see
the Appendix), with the result
    \begin{equation} \label{Ez_lossy_wall}
    E_{z}(z,t)=
    {q\over 2\pi b}\sqrt{{c\over \sigma s^{3}}},
    \end{equation}
for $s>0$. For the points where $s<0$, located in front of the
charge, $E_{z}=0$ in agreement with the causality principle. The
positive sign of $E_{z}$ indicates that the trailing charge (if it
has the same sign as $q$) will be accelerated in the wake.

In our derivation we assumed that the magnetic field on the wall is
the same as in the case of perfect conductivity. However, the
magnetic field is generated not only by the beam current, but also by
a displacement current,
    \begin{equation}
    j_{z}^{\mathrm disp}=
    {1\over 4\pi}
    \frac{\partial E_{z}}{\partial t}.
    \end{equation}
that vanishes in the limit of perfect conductivity. To be able to
neglect the corrections to $H_\theta$ due to $j_{z}^{\mathrm disp}$,
we must require the total displacement current to be much less then
the beam current. In the Fourier representation, the time derivative
$\partial /\partial t$ reduces to multiplication by $-i\omega$, and
the requirement is
    \begin{equation}
    \pi b^2 \frac{1}{4\pi}\omega E_{z\omega}
    \ll I_\omega=\frac{q}{2\pi},
    \end{equation}
or
    \begin{equation}
    \left({\omega \over c}\right)^{3/2} \ll \sqrt{{4\pi\sigma \over
    cb^{2}}}.
    \end{equation}
In the space-time domain, the inverse wavenumber $c/\omega$
corresponds to the distance $s$, and the condition of applicability
of Eq. (\ref{Ez_lossy_wall}) is,
    \begin{equation}\label{zlimitation}
    s \gg s_{0}=\left ({cb^{2} \over 4 \pi \sigma} \right)^{1/3}.
    \end{equation}

The behavior of  $E_z$ for very small values of $s$, $s < s_0$, can
be found in Ref. 6.
Here we note only that the singularity in Eq. (\ref{Ez_lossy_wall})
saturates at small $s$, and the electric field changes sign and
becomes negative at $s=0$. This field decelerates the leading charge,
as expected from the energy balance consideration.

%
%
%

\section{Wake Definition}\label{wake_definition_sec}
The electromagnetic interaction of charged particles in accelerators
with the surrounding environment is usually a relatively small effect
that can be considered as a perturbation. In the zeroth
approximation, we can assume that the beam moves with a constant
velocity along a straight line. We solve Maxwell's equation, find the
fields, and then take into account the effect of these fields on a
particle's motion. In this approach we neglect the second-order
effects because the motion along the perturbed orbit can only
slightly change the fields computed in the zeroth approximation.
Those corrections are usually small, especially for ultrarelativistic
particles.

Another important feature of the interaction between the generated
electromagnetic field and the particles is that in many cases of
practical importance it is localized in a region small compared with
the length of the beam orbit. It also occurs on a time scale much
smaller than the characteristic oscillation times of the beam in the
accelerator (such as betatron and synchrotron periods). This allows
us to consider this interaction in the impulse approximation and
characterize it by the amount of momentum transferred to the
particle.

Taking into account the above considerations, we will introduce
the notion of the \emph{wake} in the following way.
        \begin{figure}[ht]
        \begin{center}
        \includegraphics[scale=0.7]{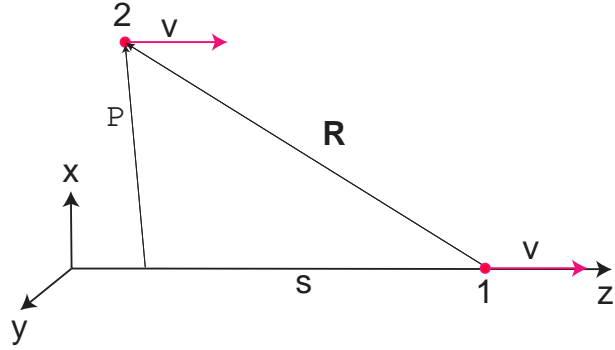}
        \end{center}
        \caption{A leading particle 1 and a trailing particle
        2 move parallel to each other in a vacuum chamber.
        \label{wake_definition1}
        }
        \end{figure}
Consider a leading particle 1 of charge $q$ moving along axis
$z$ with a velocity  close to the speed of light, $v\approx c$,
so that $z=ct$  (see Fig. \ref{wake_definition1}). A trailing
particle 2 of unit charge moves parallel to the leading one, with
the same velocity, at a distance $s$ with offset $\vec{\rho}$
relative to the $z$-axis. The vector $\vec{\rho}$ is a
two-dimensional vector perpendicular to the $z$-axis, $\vec{\rho}
= (x,y)$. Although the two particles move in a vacuum, there are
material boundaries in the problem that scatter the
electromagnetic field and result in interaction between the
particles.

Let us assume that we solved Maxwell's equation and found the
electromagnetic field generated by the first particle. We calculate
the change of the momentum $\Delta \vec{p}$ of the second particle
caused by this field  as a function of the offset $\vec{\rho}$ and
the distance $s$,
    \begin{equation}\label{deltaP}
    \Delta \vec{p}(\vec{\rho},s)
    =\int_{-\infty}^{\infty}
    dt
    \left[\vec{E}(\vec{\rho},z,t)+
    \hat{\vec{z}}\times \vec B(\vec{\rho},z,t)\right]_{z=ct-s}\,.
    \end{equation}
Note that we integrate here along a straight line --- the unperturbed
orbit of the second particle. The integration limits are extended
from minus to plus infinity, assuming that the integral converges.

Since the beam dynamics is different in the longitudinal and
transverse directions, it is useful to separate the longitudinal
momentum $\Delta p_z$ from the transverse component $\Delta
\vec{p}_\perp$. With the proper sign and the normalization factor
$c/q$, these two components are called the \emph{longitudinal} and
\emph{transverse wake functions} (or simply \emph{wakes}),
    \begin{eqnarray}\label{wakedefinitions}
    w_{l}(\vec{\rho},s)&=&
    -\frac{c}{q}\Delta p_z
    =-\frac{c}{q}\int
    dtE_{z}|_{z=ct-s},\nonumber
    \\
    \vec{w}_{t}(\vec{\rho},s)&=&
    \frac{c}{q}\Delta \vec{p}_\perp
    =
    \frac{c}{q}\int dt
    \left[\vec{E}_{\perp}+\hat{\vec{z}}\times \vec B\right]_{z=ct-s}.
    \end{eqnarray}
Note the minus sign in the definition of $w_{l}$ --- it is introduced
so that the positive longitudinal wake corresponds to the energy loss
of the trailing particle (if both the leading and trailing particles
have the same sign of charge). The defined wakes have dimension
$\cm^{-1}$ in CGS units and V/C in SI units.\footnote{A useful
relation between the units is $1~{\rm V/pC}=1.11~\cm^{-1}$.}

Because of the causality principle, the wakefield does not propagate
in front of the leading charge, hence
    \begin{equation}\label{causality}
    w_{l}(\vec{\rho},s) \equiv 0,\qquad
    \vec{w}_{t}(\vec{\rho},s) \equiv 0,\qquad
    \mathrm{for }\,\,s<0
    \,.
    \end{equation}

It was assumed above that the electromagnetic field is localized in
space and time and the integral in Eq. (\ref{deltaP}) converges.
There are cases, however, where this is not true and the source of
the wake is distributed uniformly along an extended path, such as the
resistive wall wake of a long pipe, considered in Section
\ref{res_wall_section}. In this case it is more convenient to
introduce the wake per unit length of the path by dropping the
integration in Eq. (\ref{deltaP}):
    \begin{eqnarray}\label{wakedefinitions2}
    w_{l}(\vec{\rho},s)&=&
    -\frac{1}{q}E_{z}|_{z=ct-s},\nonumber
    \\
    \vec{w}_{t}(\vec{\rho},s)&=&
    \frac{1}{q}
    \left[\vec{E}_{\perp}+\hat{\vec{z}}\times \vec B\right]_{z=ct-s}.
    \end{eqnarray}
In this definition, the wakes acquire an additional dimension of
inverse length, and has the dimension $\cm^{-2}$ in CGS and V/C/m in
SI.

\section{Panofsky-Wenzel Theorem}

Several general relations between longitudinal and transverse
wakes can be obtained from Maxwell's equation without specifying the
boundary condition for the fields.

Let us introduce the vector $\vec{R}=(\vec{\rho},-s)$ (the
negative sign in front of $s$ is due to measuring
$s$ in the negative direction of $z$, see
Fig.~\ref{wake_definition1}) and consider momentum $\Delta
\vec{p}$ in Eq. (\ref{deltaP}) as a function of $\vec{R}$. Let us
assume that the electric and magnetic fields are specified through
the vector potential $\vec{A}(\vec{r},t)$ and the scalar potential
$\phi(\vec{r},t)$, and compute $\Delta \vec{p}$ for the given
fields. It is convenient to use the Lagrangian formulation of
the equations of motion,\footnote{This approach to the derivation of
the Panofsky-Wenzel theorem is due to A. Chao.}
    \begin{equation} \label{lagr_eq}
    \frac{d}{dt}\frac{\partial L}{\partial  \vec{v}}
    =\frac{\partial L}{\partial  \vec{r}}
    \equiv \nabla L,
    \end{equation}
with the Lagrangian for the trailing \emph{unit} charge in the
electromagnetic field
    \begin{equation} \label{lagrangian}
    L=-mc^2\sqrt{1-\frac{v^2}{c^2}}+\frac{1}{c}\vec{A}\vec{v}-\phi.
    \end{equation}
Putting Eq. (\ref{lagrangian}) into Eq. (\ref{lagr_eq}) yields
($\vec{p}=m\gamma\vec{v}$)
    \begin{equation}
    \frac{d}{dt}\left(
    \vec{p}+\frac{1}{c}\vec{A}
    \right)
    =
    \nabla
    \left(
    \frac{1}{c}\vec{A}\vec{v}-\phi
    \right).
    \end{equation}
Now, integrating this equation along the orbit of the trailing
particle, $x=\const$, $y=\const$ and $z=ct-s$, and assuming that the
fields $\vec{A}$ and $\phi$ vanish at infinity, we find
    \bea
    \Delta \vec{p}(\vec{R})&=&
    \int dt
    \nabla
    \left(
    \frac{1}{c}\vec{A}\vec{v}-\phi
    \right)
    \nonumber\\
    &=&
    \frac{q}{c}\nabla_{\vec{R}} W(\vec{R}),
    \eea
where we introduced the \emph{wake potential} $W$,
    \bea
     W(\vec{R})&=&
     \frac{c}{q}
    \int dt
    \left(
    \frac{1}{c}\vec{A}\vec{v}-\phi
    \right)
    \nonumber\\
    &=&
    \frac{c}{q}\int dt
    \left(
    A_z-\phi
    \right).
    \eea
In the last equation we used $\vec{v}\approx c\hat{\vec{z}}$.

We just proved an important relation that states that all three
components of the vector $\Delta\vec{p}$ can be obtained by
differentiation of a single scalar function $W$. Recalling now the
relation between the components of $\Delta\vec{p}$ and the wakes, Eq.
(\ref{wakedefinitions}), we find that
    \begin{equation}\label{wakes_via_W}
    w_{l} = -\frac{\pa W}{\pa (-s)}=\frac{\pa W}{\pa s},
    \qquad
    \vec{w}_{t} = \nabla_{\vec{\rho}}W,
    \end{equation}
and hence
    \begin{equation}\label{PW2}
    {\partial \vec{w}_{t} \over \partial s}=\nabla_{\vec{\rho}}w_{l}.
    \end{equation}
This relation is usually referred to as the Panofsky-Wenzel theorem.
Note that $\nabla_{\vec{\rho}}$ is a two-dimensional gradient with
respect to coordinates $x$ and $y$.

One of the most important computational applications of the
Panofsky-Wenzel theorem is that knowledge of the longitudinal
wake function $w_l$ allows us to find the transverse wake $w_t$ by
means of a simple integration of Eq. (\ref{PW2}).

We now prove another important property of $W$: it is a harmonic
function of variables $x$ and $y$,
    \begin{equation} \label{W_laplace}
    \Delta_\perp W \equiv
    \frac{\pa^2 W}{\pa x^2}+\frac{\pa^2 W}{\pa y^2}
    =0.
    \end{equation}
To prove this, we will use the fact that both $\vec{A}$ and $\phi$
satisfy the wave equation in free space, $(\pa^2/\pa
t^2-c^{2}\Delta)\vec{A}=(\pa^2/\pa t^2-c^{2}\Delta)\phi=0$. Hence
 \bea
    0 &=&
    \frac{c}{q}\int dt
    \left(\frac{\pa^2}{\pa t^2}-c^{2}\Delta\right)
    \left(A_z-\phi\right)
    \nonumber\\
    &=&
    -\frac{c}{q}\int dt
    \left(\frac{\pa^2 }{\pa x^2}+\frac{\pa^2 }{\pa y^2}\right)
    \left(
    A_z-\phi
    \right)
    +
    \frac{c}{q}\int dt
    \left(\frac{\pa^2 }{\pa t^2} -c^{2}\frac{\pa^2}{\pa z^2}\right)
    \left(
    A_z-\phi
    \right)
    \nonumber\\
    &=&
    -\frac{\pa^2 W}{\pa x^2}-\frac{\pa^2 W}{\pa y^2}
    +
    \frac{c}{q}\int dt
    \left(\frac{\pa }{\pa t} +c\frac{\pa}{\pa z}\right)
    \left(\frac{\pa }{\pa t} -c\frac{\pa}{\pa z}\right)
    \left(A_z-\phi\right).
    \eea
The last integral in this equation vanishes because
    \begin{equation}
    \frac{\pa }{\pa t} +c\frac{\pa}{\pa z}
    \approx
    \frac{\pa }{\pa t} +\vec{v}\nabla
    =
    \frac{d}{dt}
    \end{equation}
and
 \bea
    &&
    \int dt
    \left(\frac{\pa }{\pa t} +c\frac{\pa}{\pa z}\right)
    \left(\frac{\pa }{\pa t} -c\frac{\pa}{\pa z}\right)
    \left(A_z-\phi\right)
    \nonumber\\
    &=&
    \int dt
    \frac{d}{dt}
    \left(\frac{\pa }{\pa t} -c\frac{\pa}{\pa z}\right)
    \left(A_z-\phi\right)
    \nonumber \\
    &=&
    0.
    \eea

It is interesting that the wake potential $W$ turns out to be a
relativistic invariant. A covariant expression for it can be written
as
    \begin{equation}
    W =-\frac{1}{q}
    \int_{-\infty}^{\infty}A_k u^k d\tau,
    \end{equation}
where $A_k=(\phi,-\vec{A})$ is the 4-vector potential,
$u^k=(c\gamma,c\gamma\vec{v})$ is the 4-vector velocity, and $\tau$ is
the proper time for the particle.

\section{Systems with a Symmetry Axis}

\begin{figure}[ht]
        \begin{center}
        \includegraphics[scale=0.7]{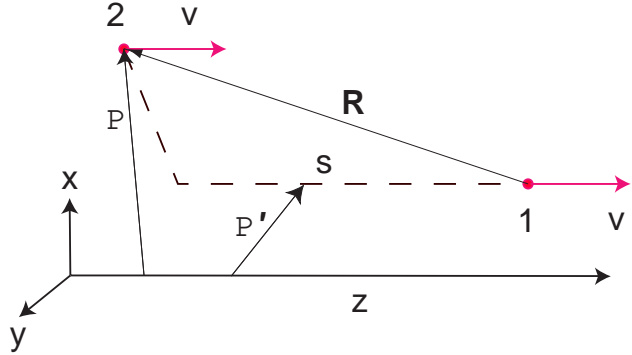}
        \end{center}
        \caption{Both the leading particle 1 and the trailing particle
        2 are offset relative to the axis of the chamber.
        \label{wake_definition}
        }
        \end{figure}

In Section \ref{wake_definition_sec} we defined the wake as a
function of the trailing particle offset relative to the path of
the leading particle. In practical applications we are also
interested in how the wake depends on the trajectory of the
leading particle. We will assume that the system under
consideration has a symmetry axis, and choose it as the $z$-axis
of the coordinate system (see Fig.~\ref{wake_definition}).
Now the leading particle 1 moves in the $z$ direction with an offset
given by vector $\vec{\rho'}$, and the trailing particle travels
parallel to the leading one, with the same velocity, at a distance
$s$ behind the leading one, and with offset $\vec{\rho}$ relative to
the axis. The vectors $\vec{\rho'}$ and $\vec{\rho}$ are the
two-dimensional vectors perpendicular to the $z$-axis. The wake is
still defined by Eq. (\ref{wakedefinitions}), but now it will be
considered as a function of $\vec{\rho}'$, $\vec{\rho}$, and $s$
    \bea
    w_{l}&=&
    w_{l}(\vec{\rho},\vec{\rho}',s),\nonumber
    \\
    \vec{w}_{t}&=&
    \vec{w}_{t}(\vec{\rho},\vec{\rho}',s).
    \eea

Usually the vacuum chamber is designed so that the system axis serves
as an ideal orbit for the beam. Deviations from it are relatively
small, and both vectors $\vec{\rho}$ and $\vec{\rho}'$ are typically
much smaller than the size of the vacuum chamber. We can neglect them
in $w_{l}$ and introduce the longitudinal wake function that depends
only on $s$,
    \begin{equation}\label{longwake}
    w_{l}(s)=w_{l}(0,0,s).
    \end{equation}

If the vacuum chamber also has some symmetry elements (e.g., it has
either circular, elliptical or rectangular cross section), the
transverse wake on the axis, where $\vec{\rho},\vec{\rho}' =0$,
vanishes, $\vec{w}_{t}(0,0,s)=0$. For small values of
$\vec{\rho},\vec{\rho}'$ we can expand
$\vec{w}_{t}(\vec{\rho},\vec{\rho}',s)$ keeping only the lowest-order
linear terms. That gives a tensor relation between the transverse
wake and the offsets,
    \begin{equation}
    \vec{w}_{t}(\vec{\rho},\vec{\rho}',s)
    =\overset{\leftrightarrow}{\vec{W}}_1(s) \vec{\rho}
    +\overset{\leftrightarrow}{\vec{W}}_2(s) \vec{\rho'},
    \end{equation}
where $\overset{\leftrightarrow}{\vec{W}}_1$ and
$\overset{\leftrightarrow}{\vec{W}}_2$ are the two-dimensional
tensors of rank 2. An example of the wake calculation for elliptical
and rectangular cross sections of the pipe can be found in Ref. 7.

\section{Axisymmetric Systems}

\begin{figure}[ht]
        \begin{center}
        \includegraphics[scale=0.7]{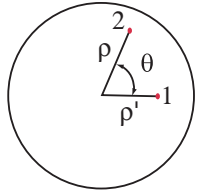}
        \end{center}
        \caption{Vectors $\vec{\rho}$ and $\vec{\rho}'$ in
        an axisymmetric system.
        \label{axisymmetry}
        }
        \end{figure}

In an axisymmetric system $W$ depends only on the absolute values of
$\rho$, $\rho'$, and the angle $\theta$ between them. We can always
chose a coordinate system such that the vector $\vec{\rho'}$ lies
in the $x$-$z$ plane (see Fig. \ref{axisymmetry}), so that the
potential function $W$ will be a periodic even function of angle
$\theta$ in a cylindrical coordinate system.
Decomposing $W$ in Fourier series in $\theta$ yields
    \begin{equation}
    W(\rho,\rho',\theta,s)=\sum_{m=0}^\infty
    W_m(\rho,\rho',s)\cos m\theta.
    \end{equation}
Putting this equation into Eq. (\ref{W_laplace}) gives
    \begin{equation}\label{Wm_eq}
    \sum_{m=0}^\infty
    \left(
    \frac{1}{\rho}
    \frac{\pa}{\pa \rho}
    \rho
    \frac{\pa W_m}{\pa \rho}
    -
    \frac{m^2}{\rho^2} W_m
    \right)
    \cos m\theta
    =0,
    \end{equation}
from which we can find an explicit dependence of $W_m$ of $\rho$,
    \begin{equation}\label{Wm}
    W_m(\rho,\rho',s)= A_m(\rho',s){\rho}^m.
    \end{equation}
In Eq. (\ref{Wm}) we discarded a singular solution of Eq.
(\ref{Wm_eq}) $W_m \propto \rho^{-m}$.

It is also possible to find
the dependence of $W_m$ versus $\rho'$ (see \cite{bane85ww}), which
turns out to be
    \begin{equation}
    A_m(\rho',s)= F_m(s)(\rho')^m.
    \end{equation}
Using Eq. (\ref{wakes_via_W}) we now find for the longitudinal and
transverse wake functions
    \begin{equation}
    w_{l}=\sum  w_{l}^{(m)}, \qquad \vec{w}_{t}=\sum \vec{w}_{t}^{(m)}
    \end{equation}
where
    \begin{eqnarray}\label{axisymsystems1}
    w_{l}^{(m)}&=&(\rho')^{m}\rho^{m}F_m'(s)\cos m\theta,\nonumber
    \\
    \vec{w}_{t}^{(m)}&=&m(\rho')^{m}\rho^{m-1}F_m(s)\left[\hat{\vec{r}}\cos
    m\theta- \hat{\vec{\theta}}\sin m\theta \right],
    \end{eqnarray}
where  $\hat{\vec{r}}$ and $\hat {\vec{\theta}}$ are the unit vectors
in the radial and azimuthal directions in the cylindrical coordinate
system. Remember that in this equation we assume that the leading
particle is in the plane $\theta=0$.

Equations (\ref{axisymsystems1}) are valid for arbitrary values of
$\rho$ and $\rho'$. Near the axis, where the offsets are small, the
higher-order terms with large values of $m$ in these equations also
become small. In this case, we can keep only the lower-order terms
with $m=0$ (\emph{monopole}) and $m=1$ (\emph{dipole}) wakes. For the
monopole wake we find
    \begin{equation}\label{long_wake_pot}
    w_{l} \equiv w_{l}^{(0)}= F_0'(s),
    \end{equation}
which shows that the longitudinal wake does not depend on the radius
in an axisymmetric system. We also see that the monopole transverse
wake vanishes, $w_t^{(0)}=0$.

Since $w_{l}^{(0)}$ does not depend on $\rho'$, sometimes it is
more convenient to compute the monopole wake for an offset orbit,
$\rho' \ne 0$, rather than on the axis.

    \begin{figure}[ht]
    \begin{center}
    \includegraphics[scale=0.7]{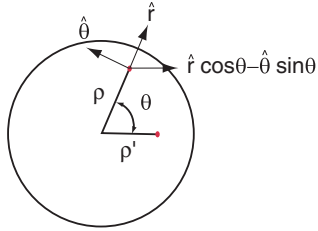}
    \end{center}
    \caption{Vectors $\vec{\rho}$ and $\vec{\rho}'$ and unit vectors
    in cylindrical coordinate system.
    \label{axisymmetry1}
    }
    \end{figure}

For the dipole wake ($m=1$),  the vector $\hat{\vec{r}}\cos
\theta-\hat{\vec{\theta}}\sin \theta $ lies in the direction of the
$x$ axis, that is in the direction of ${\vec{\rho}'}$, see Fig.
\ref{axisymmetry1}. Hence,
    \begin{equation}\label{axisymsystems2}
    \vec{w}_{t}^{(1)}={\vec{\rho}'}F(s).
    \end{equation}
The wake given by Eq. (\ref{axisymsystems2}) is usually normalized by
the absolute value of the offset $\rho'$, and the scalar function
${w}_{t}^{(1)}/\rho'$ is called the transverse dipole wake $w_t$,
    \begin{equation}\label{trans_wake_def}
    w_t(s)=F(s).
    \end{equation}
Such a transverse wake has the dimension $\cm^{-2}$ or
V/C/m.\footnote{If the original wake is defined per unit length, as
in Eq. (\ref{wakedefinitions2}), then $w_t$ will have the dimension
V/C/m$^2$ or $\cm^{-3}$.} In this definition, a positive transverse
wake means a kick in the direction of the offset of the driving
particle (if both particles have the same charge).

\section{Resistive Wall Wake Functions}\label{resist_wall_wakes}

We are now in a position to find the wakes generated by a particle in
a circular pipe with resistive walls. Using Eq. (\ref{Ez_lossy_wall})
and the definition Eq. (\ref{wakedefinitions}) gives the longitudinal
wake
    \begin{equation}\label{resistive_wall_long_wake}
    w_l(s)=
    -{1\over 2\pi b}\sqrt{{c\over \sigma s^{3}}}\,.
    \end{equation}
The minus sign here means that the trailing charge is accelerated in
the wakefield.

Let us now calculate the dipole transverse wake due to the resistive
wall. First, we need to solve for the electromagnetic field of the
leading charge $q$ moving with an offset $\vec{\rho}'$ in a circular
pipe. From the point of view of excitation of dipole modes, this
charge can be considered as having a dipole moment
$\vec{d}=q\vec{\rho}'$. In the zeroth approximation of perturbation
theory, the electromagnetic field of a dipole moving with the speed
of light in a \emph{perfectly conducting} pipe is
    \begin{equation} \label{dipole_field}
    \vec{E}=2\delta(z-ct)
    \left[
    \frac{2(\vec{d}\cdot\vec{r})\vec{r}-\vec{d} r^2}{r^{4}}
    +
    \frac{\vec{d}}{b^2}
    \right]
    \,,
    \qquad
    \vec{H}=\vec{\hat{z}}\times\vec{E}\,.
    \end{equation}
The first term in the expression for $\vec{E}$ is a vacuum field of a
relativistic dipole, and the second one is due to the image charges,
which are generated in order to satisfy the boundary condition on the
metal surface.

Following the derivation in Section \ref{res_wall_section},
we find the magnetic field on the wall,
    \begin{equation}
    H_\theta = \frac{4d}{b^2}\cos\theta\delta(z-ct),
    \end{equation}
and take its Fourier transform,
    \begin{equation}
    H_{\theta\omega} = \frac{2d}{\pi c b^2}\cos\theta\,,
    \end{equation}
where the angle $\theta$ is measured from the direction of $\vec{d}$.
Then using the Leontovich boundary condition, Eq. (\ref{Ez}), for the
electric field $E_{z\omega}$,
    \begin{equation}
    E_{z\omega}|_{r=b}
    =-(1-i)\sqrt{\frac{\omega}{8\pi
    \sigma}}
    \frac{2d}{\pi c b^2}\cos\theta,
    \end{equation}
and making the inverse Fourier transformation, we obtain $E_z$ on the
wall
    \begin{equation}
    E_{z}(z,\rho=b,\rho',t)=
    {q\rho'\over \pi b^2}\sqrt{{c\over \sigma s^{3}}}
    \cos\theta\,,
    \end{equation}
where $s=ct-z$. Recalling that, according to Eq.
(\ref{axisymsystems1}) the dipole wake is a linear function of
$\rho$, we conclude that
    \begin{equation} \label{Ez_lossy_wall_dipole}
    E_{z}(z,\rho,\rho',t)=
    {q\rho\rho'\over \pi b^3}\sqrt{{c\over \sigma s^{3}}}
    \cos\theta\,,
    \end{equation}
and the function $F_1'(s)$ is
    \begin{equation}
    F_1'(s)=
    -{1\over \pi b^3}\sqrt{{c\over \sigma s^{3}}}
    \,,
    \end{equation}
which gives the following result for the transverse wake defined by
Eq. (\ref{trans_wake_def}):
    \begin{equation}
    w_t(s)=
    {2\over \pi b^3}\sqrt{{c\over \sigma s}}
    \,.
    \end{equation}
Analogous to the longitudinal wake, Eq. (\ref{Ez_lossy_wall}), this
formula is valid only for $s \gg s_0$ (see Eq. (\ref{zlimitation})).

\section{Wakefield in a Bunch of Particles}

Up to now we have studied the interaction of two point charges
traveling some distance $s$ apart. If a beam consists of $N$
particles with the distribution function $\lambda(s)$ (defined so
that $\lambda (s)ds$ gives the probability of finding a particle near
point $s$), a given particle will interact with all other
particles of the beam. To find the change of the longitudinal
momentum of the particle at point $s$ we need to sum the wakes from
all other particles in the bunch,
    \begin{equation}\label{momentum_change}
    \Delta p_z(s)
    =\frac{Ne^2}{c}\int_{s}^{\infty}
    ds' \lambda(s')w_l(s'-s).
    \end{equation}
Here we use the causality principle and integrate only over the part
of the bunch ahead of point $s$. In the ultrarelativistic limit the
energy change $\Delta E(s)$ caused by the wakefield is equal to
$c\Delta p_z$, so Eq. (\ref{momentum_change}) can also be rewritten
as
    \begin{equation}\label{energy_change}
    \Delta E(s)
    =Ne^2\int_{s}^{\infty}
    ds' \lambda(s')w_l(s'-s).
    \end{equation}

Two integral characteristics of the strength of the wake are the
average value of the energy loss $\Delta E_{\mathrm av}$, and the rms
spread in energy generated by the wake $\Delta E_\rms$. These two
quantities are defined by the following equations:
    \begin{equation}\label{average_energy_change}
    \Delta E_{\mathrm av}
    =
    \int_{-\infty}^{\infty}
    ds\Delta E(s) \lambda(s)
    ,
    \end{equation}
and
    \begin{equation}\label{RMS_energy_change}
    \Delta E_\rms
    =
    \left[\int_{-\infty}^{\infty}
    ds(\Delta E(s)-{\Delta E_{\mathrm av}})^2 \lambda(s)
    \right]^{1/2}
    .
    \end{equation}

As an example, let us calculate ${\Delta E_{\mathrm av}}$ and $\Delta
E_\rms$ for the resistive wall wake given by Eq.
(\ref{resistive_wall_long_wake}) and a Gaussian distribution
function,
    \begin{equation}
    \rho(s)=
    \frac{1}{\sqrt{2\pi}\sigma_s}
    \exp
    \left(-\frac{s^2}{2\sigma_s^2}\right),
    \end{equation}
where  $\sigma_s$ is the rms bunch length. Note that, since $w$ in Eq.
(\ref{resistive_wall_long_wake}) is the wake per unit length of the
pipe, we need to multiply the final answer by the pipe length $L$.

\begin{figure}[ht]
        \begin{center}
        \includegraphics[scale=0.8]{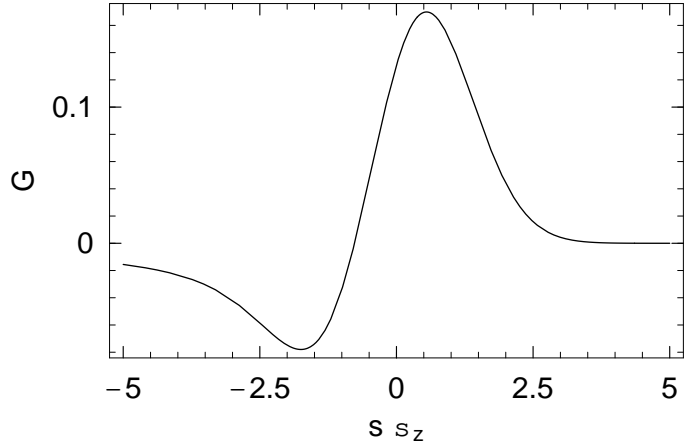}
        \end{center}
        \caption{Plot of the function $G\left({s}/{\sigma_z}\right)$.
        \label{long_wake_fig}
        }
        \end{figure}

A direction substitution of the wake Eq.
(\ref{resistive_wall_long_wake}) into Eq. (\ref{energy_change}) gives
a divergent integral when $s'\rightarrow s$.\footnote{The integral
diverges because Eq. (\ref{resistive_wall_long_wake}) is not valid
for very small values of $s$, see Eq. (\ref{zlimitation}).} To remove
the divergence, we need to recall that according to Eq.
(\ref{long_wake_pot}) the longitudinal wake is equal to the
derivative of the longitudinal wake potential, $w_l=F'_0(s)$ with
$F_0=(\pi b)^{-1}\sqrt{c/\sigma s}$ for $s>0$, and $F_0=0$ for $s<0$.
We then rewrite Eq. (\ref{energy_change}) as
    \bea\label{energy_change_thru_F}
    \Delta E(s)
    &=&
    Ne^2L\int_{-\infty}^{\infty}
    ds' \lambda(s')\frac{dF_0(s'-s)}{ds}
    \nonumber \\
    &=&
    -Ne^2L\int_{s}^{\infty}
    ds' \frac{d\lambda(s')}{ds}F_0(s'-s)
    \nonumber \\
    &=&{Ne^2 Lc^{1/2} \over b \sigma_z^{3/2}\sigma^{1/2}}
    G\left(\frac{s}{\sigma_z}\right),
    \eea
where the function $G(x)$ is
    \begin{equation}\label{G}
    G(x)=\frac{1}{2^{1/2}\pi^{3/2}}\int_x^\infty
    \frac{ye^{-y^2/2}dy}{\sqrt{y-x}}.
    \end{equation}
The plot of the function $G\left({s}/{\sigma_z}\right)$ is shown in Fig.
\ref{long_wake_fig}, where the positive values of $s$ correspond to
the head of the bunch. We see that in the resistive wake the
particles lose energy in the head of the bunch and get accelerated in
the tail. On the average, of course, the losses overcome the gain.
For the average energy loss one can find an analytical result:
    \begin{equation}
    \Delta E_{\mathrm av} =
    {\Gamma({3\over 4}) \over 2^{3/2} \pi^{3/2}}
    {Ne^2 c^{1/2} \over b \sigma_z^{3/2}\sigma^{1/2}}.
    \end{equation}
Numerical integration of Eq. (\ref{RMS_energy_change}) shows that the
energy spread generated by the resistive wake is approximately equal
to $\Delta E_{\mathrm av}$:
    \begin{equation}
    \Delta E_\rms =1.06\, \Delta E_{\mathrm av}.
    \end{equation}

\begin{figure}[ht]
    \begin{center}
    \includegraphics[scale=0.8]{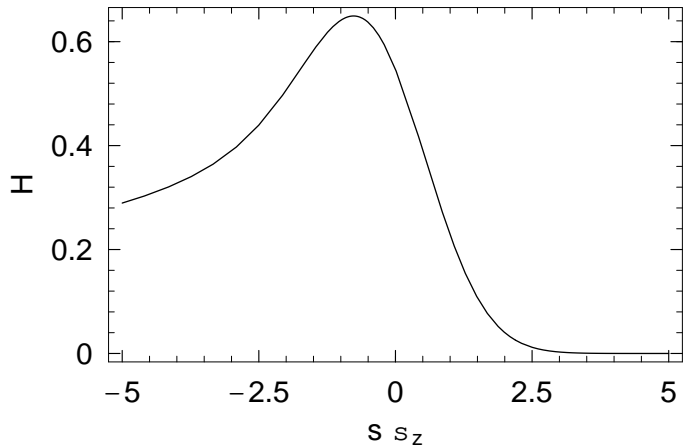}
    \end{center}
    \caption{Plot of the function $H\left({s}/{\sigma_z}\right)$.
    \label{tran_wake_fig}}
    \end{figure}

If the beam is traveling in the pipe with an offset $y$ relative to
the axis, it will be deflected in the direction of the offset, by the
transverse wakefields. To calculate the deflection angle $\theta$ we
use the relation
    \bea
    \theta(s)&=&\frac{\Delta p_\perp(s)}{p}
    =yL\frac{Ne^2}{cp}\int_{s}^{\infty}
    ds' \lambda(s')w_t(s'-s)
    \nonumber \\
    &=&{NLr_eyc^{1/2} \over \gamma b^3
    \sigma_z^{1/2}\sigma^{1/2}}
    H\left(\frac{s}{\sigma_z}\right),
    \eea
where the function $H(x)$ is
    \begin{equation}
    H(x)=\frac{2^{1/2}}{\pi^{3/2}}\int_x^\infty
    \frac{e^{-y^2/2}dy}{\sqrt{y-x}}.
    \end{equation}
The plot of the function $H\left({s}/{\sigma_z}\right)$ is shown in Fig.
\ref{tran_wake_fig}.

The deflection angle averaged over the distribution function is
    \begin{equation}
    \theta_{\mathrm av} =
    {\Gamma({1\over 4}) \over 2^{1/2} \pi^{3/2}}
    {NLr_eyc^{1/2} \over \gamma b^3 \sigma_z^{1/2}\sigma^{1/2}},
    \end{equation}
and the rms spread is
    \begin{equation}
    \langle (\theta-\theta_{\mathrm av})^2 \rangle^{1/2} =
    A^{1/2}
    {NLr_eyc^{1/2} \over \gamma b^3 \sigma_z^{1/2}\sigma^{1/2}},
    \end{equation}
where
    \begin{equation} \label{const_A}
    A=\frac{2}{\pi^{5/2}}
    \left[
    K\left(\frac{3}{4}\right)
    -\frac{\Gamma^2(1/4)}{4\sqrt{\pi}}
    \right]
    \end{equation}
and $K$ is the complete elliptic integral. The numerical value of $
A^{1/2}$ is 0.186.

\section{Definition of Impedance and Relation Between Impedance and Wake}

Knowledge of the longitudinal and transverse wake functions gives
complete information about the electromagnetic interaction of the
beam with its environment. However, in many cases, especially in the
study of beam instabilities, it is more convenient to use the Fourier
transform of the wake functions, or \emph{impedances}. Also, it is
often easier to calculate the impedance for a given geometry of the
beam pipe, rather than the wake function. Recall, that in Section
\ref{res_wall_section} we actually first computed the Fourier
components of the wakes, and then, using the inverse Fourier
transformation, found the wakes.

For historical reasons the longitudinal $Z_{l}$ and transverse
$Z_{t}$ impedances are defined as Fourier transforms of wakes with
different factors,
    \begin{eqnarray}\label{impedancefinitions}
    Z_{l}(\omega)&=&{1\over c}\int_{0}^{\infty}
    dsw_{l}(s) e^{i\omega s/c}\,,\nonumber
    \\
    Z_{t}(\omega)&=&-{i\over c}\int_{0}^{\infty}
    dsw_{t}(s) e^{i\omega s/c}\,.
    \end{eqnarray}
Note that the integration in Eqs. (\ref{impedancefinitions}) can
actually be extended into the region of negative values of $s$,
because $w_l$ and $w_t$ are equal to zero in that region.

Impedance can also be defined for complex values of $\omega$ such
that $\Im \omega > 0$ and the integrals, Eq.
(\ref{impedancefinitions}), converge. So defined, the impedance is an
analytic function in the upper half-plane of the complex variable
$\omega$.

We must keep in mind that other authors sometimes introduce
definitions of the impedance that differ from the one given
above. In Refs. 2 and 9
the longitudinal impedance is
defined as a complex conjugate to the one given by Eq.
(\ref{impedancefinitions}). Here we follow the definitions of Refs. 1
and 10.

From the definitions in Eq. (\ref{impedancefinitions}) it follows
that the impedance satisfies the following symmetry conditions:
    \bea
    \Re Z_{l}(\omega)&=&\Re Z_{l}(-\omega),\qquad
    \,\,\,\,\Im Z_{l}(\omega)=-\Im Z_{l}(-\omega),
    \nonumber \\
    \Re Z_{t}(\omega)&=&-\Re Z_{t}(-\omega),\qquad
    \Im Z_{t}(\omega)=\Im Z_{t}(-\omega).
    \eea

The inverse Fourier transform relates the wakes to the
impedances:
    \begin{eqnarray}\label{wake_through_impedance}
    w_{l}(s)&=&{1\over 2\pi}\int_{-\infty}^{\infty}
    d\omega Z_{l}(\omega) e^{-i\omega s/c}\,,\nonumber
    \\
    w_{t}(s)&=&{i\over 2\pi}\int_{\infty}^{\infty}
    d\omega Z_{t}(\omega) e^{-i\omega s/c}\,.
    \end{eqnarray}

It turns out that the wakefield can actually be found if only the
real part of the impedance is known. Indeed, we can rewrite Eq.
(\ref{wake_through_impedance}) for $w_l$ as
    \begin{equation}\label{wake_through_impedance1}
    w_{l}(s)
    ={1\over 2\pi}\int_{\infty}^{\infty}
    d\omega
    \left[
    \Re Z_{l}(\omega) \cos\frac{\omega s}{c}
    -\Im Z_{l}(\omega) \sin\frac{\omega s}{c}
    \right]
    \,.
    \end{equation}
For negative values of $s$ this formula should give $w_l=0$, hence
    \begin{equation}\label{wake_through_impedance2}
    0
    =\int_{\infty}^{\infty}
    d\omega
    \left[
    \Re Z_{l}(\omega) \cos\frac{\omega s}{c}
    +\Im Z_{l}(\omega) \sin\frac{\omega |s|}{c}
    \right],
    \end{equation}
from which it follows that
   \begin{equation}\label{wake_through_impedance3}
    w_{l}(s)
    ={1\over \pi}\int_{\infty}^{\infty}
    d\omega
    \Re Z_{l}(\omega) \cos\frac{\omega s}{c}
    ={2\over \pi}\int_{0}^{\infty}
    d\omega
    \Re Z_{l}(\omega) \cos\frac{\omega s}{c}
    \,.
    \end{equation}

A similar derivation for the transverse wake gives
   \begin{equation}\label{wake_through_impedance4}
    w_{t}(s)
    ={2\over \pi}\int_{0}^{\infty}
    d\omega
    \Re Z_{t}(\omega) \sin\frac{\omega s}{c}
    \,.
    \end{equation}

\section{Energy Loss and $\Re Z_l$}

We can relate the energy loss by the bunch to the real part of the
longitudinal impedance. Indeed,
    \bea
    \Delta E_{\mathrm av}
    &=&
    N^2e^2
    \int_{-\infty}^{\infty}
    ds \lambda(s)
    \int_{-\infty}^{\infty}
    ds' \lambda(s')w_l(s'-s)
    \nonumber \\
    &=&
    N^2e^2
    \int_{-\infty}^{\infty}
    ds \lambda(s)
    \int_{-\infty}^{\infty}
    ds' \lambda(s')
    \frac{1}{2\pi}
    \int_{-\infty}^{\infty}
    d\omega Z_l(\omega)
    e^{-i\omega(s'-s)/c}
    \nonumber\\
    &=&
    \frac{N^2e^2}{2\pi}
    \int_{-\infty}^{\infty}
    d\omega Z_l(\omega)
    |\hat{\lambda}(\omega)|^2,
    \eea
where $\hat{\lambda}(\omega)=\int_{-\infty}^{\infty} ds \lambda(s)e^{i\omega
s/c}$. Since $\hat{\lambda}(-\omega)=\hat{\lambda}^*(\omega)$, $
|\hat{\lambda}(\omega)|^2$ is an even function of $\omega$, and
    \begin{equation}
    \Delta E_{\mathrm av}
    =
    \frac{Q^2}{\pi}
    \int_{0}^{\infty}
    d\omega \Re Z_l(\omega)
    |\hat{\lambda}(\omega)|^2,
    \end{equation}
where $Q=Ne$.

For a point charge, $\lambda(s)=\delta(s)$,
$\hat{\lambda}(\omega)=1$, and the energy loss is
    \begin{equation}
    \Delta E_{\mathrm av}
    =
    \frac{e^2}{\pi}
    \int_{0}^{\infty}
    d\omega \Re Z_l(\omega).
    \end{equation}

\section{Kramers-Kronig Relations}

Equations (\ref{wake_through_impedance3}) and
(\ref{wake_through_impedance4}) relate $\Re Z(\omega)$ to the wake
function. Since $Z(\omega)$ is given by Fourier transformation of
$w(s)$, the knowledge of $\Re Z(\omega)$ allows us to find
$Z(\omega)$, and hence $\Im Z(\omega)$. This means that $\Im
Z(\omega)$ and $\Re Z(\omega)$ are functionally related to each
other. Mathematically this relation is manifested in the
Kramers-Kronig {\em dispersion} relation, which can be written as
    \begin{equation}\label{disprelation1}
    Z(\omega)=-\frac{i}{\pi}{\rm P.V.}\int_{-\infty}^{\infty}{Z(\omega')
    \over \omega'-\omega}d\omega',
    \end{equation}
where P.V. stands for the principal value of the integral. Taking the
real and imaginary parts of this equation gives explicit relations
between $\Re Z$ and $\Im Z$:
    \bea\label{disprelation2-1}
    \Re Z(\omega)&=&\frac{1}{\pi}{\rm P.V.}\int_{-\infty}^{\infty}{\Im Z(\omega')
    \over \omega'-\omega}d\omega',
    \nonumber \\
    \Im Z(\omega)&=&-\frac{1}{\pi}{\rm P.V.}\int_{-\infty}^{\infty}{\Re Z(\omega')
    \over \omega'-\omega}d\omega'.
    \eea

\section{Useful Formula for Impedance Calculation}

Assume that we have a solution of an electromagnetic problem
corresponding to the current on the axis of a pipe with the time and
space dependence given by $e^{-i\omega t+i\omega {z / c}}$.
Specifically, we know the electric field on the axis,
$E_{z\omega}(z)e^{-i\omega t}$. How can longitudinal impedance be
found in terms of $E_{z\omega}(z)$?

The longitudinal wake is equal to the integrated field $E_z$
generated by a point charge moving with the speed of light. The
current corresponding to this point charge can be decomposed into a
Fourier integral:
    \begin{equation}
    cq\delta (z-ct) = \int_{-\infty}^{\infty}d\omega
    {q \over 2\pi}
    e^{-i\omega(t-z/c)}.
    \end{equation}
Since we know the electric field generated by each harmonic, we can
find the field due to the point charge as a superposition of
$E_{z\omega}$:
    \begin{equation}
    E_{z}(z,t)=
    {q \over 2\pi}
    \int_{-\infty}^{\infty}d\omega
    E_{z\omega}(z)
    e^{-i\omega t}.
    \end{equation}
For $w_{l}$ we then have
    \bea\label{longimped1}
    w_{l}(s)&=&-\frac{1}{2\pi}\int_{-\infty}^\infty
    d\omega d(ct)E_{z\omega}(ct-s)e^{-i\omega t}\nonumber
    \\
    &=&-\frac{1}{2\pi}\int_{-\infty}^\infty
    d\omega dzE_{z\omega}(z)e^{-i\omega (z+s)/c}.
    \eea
Comparing Eq. (\ref{longimped1}) with Eq.
(\ref{impedancefinitions}) we find
    \begin{equation}\label{impedformula1}
    Z_{l}(\omega)=
    -\int_{-\infty}^{\infty}
    dzE_{\omega}(z)e^{-i\omega z/c}.
    \end{equation}
Hence the longitudinal impedance can be obtained simply by making
Fourier transformation of $E_{z\omega}(z)$.

\section{Small Pillbox Cavity in Round Pipe}

\begin{figure}[ht]
 \begin{center}
 \includegraphics[scale=0.8]{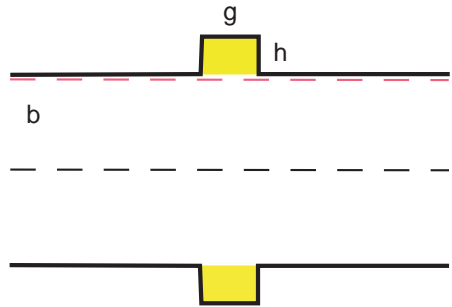}
 \end{center}
 \caption{Small pillbox cavity in a round pipe.
 The dashed line near the wall shows the integration path in Eq. (\ref{eq91}).
 \label{pillbox}}
 \end{figure}

As an example of using Eq. (\ref{impedformula1}) we will derive here
the longitudinal impedance $Z_l(\omega)$ for a small axisymmetric
cavity (pillbox) in a round perfectly conducting pipe, see Fig.
\ref{pillbox}. We assume that the wavelength associated with the
frequency $\omega$ is much larger than the dimension of the pillbox,
$c/\omega \gg g,h$.

First, we need to find the solution of Maxwell's equations
corresponding to a unit current  $I_\omega = e^{-i\omega t+i\omega {z /
c}}$ on the axis. Since the cavity is small, the magnitude of the
magnetic field at the location of the cavity ($z=0$) is approximately equal
to $H_\theta$ on the wall of the pipe
    \begin{equation}
    H_\theta =\frac{2}{bc}.
    \end{equation}
Because $E_{z\omega}$ does not depend on $r$ (see Section
\ref{res_wall_section}) we can choose the integration path in Eq.
(\ref{impedformula1}) close to the wall, as shown in Fig.
\ref{pillbox}, rather than the pipe axis. Along this path
$E_{z\omega}$ is not equal to zero only in the cavity gap, where
$|z|\sim h$, and $e^{-i\omega z}\approx 1$. We have
    \bea
    Z_{l}(\omega)&=&
    -\int_{-\infty}^\infty
    dzE_{z\omega}(z)e^{-i\omega z/c}
    \nonumber \\
    &\approx&
        -\int_{-\infty}^\infty
    dzE_{z\omega}(z)
    =
    \frac{1}{c}\frac{d\Phi}{dt}
    =-\frac{i\omega}{c} \Phi,
    \eea
where $\Phi$ is the magnetic flux in the cross section of the cavity, $\Phi = H_\theta
hg$. As a result,
    \begin{equation}\label{eq91}
    Z_{l}(\omega)
    =-i\omega \frac{2gh}{bc^2}
    =-i\omega Z_0\frac{gh}{2\pi bc},
    \end{equation}
where $Z_0=4\pi/c=377$ Ohm. What we obtained is a purely
\emph{inductive} longitudinal impedance, which can be rewritten as
    \begin{equation}\label{inductive_imped}
    Z_{l}(\omega)=-i\frac{\omega}{c^2}{\cal L},
    \end{equation}
where the inductance ${\cal L}={2gh}/{b}$. In CGS units the
inductance has a dimension of cm, 1 cm = 1 nH.

A more detailed calculation \cite{kurennoy94s} shows that our method
gives only an approximate solution of the problem. In addition to the
solenoidal electric field generated by the time-varying magnetic flux
in the cavity, there is also a contribution due to the potential
component of the electric field. This results in a different
numerical coefficient in Eq. (\ref{eq91}) which depends on the ratio
$g/h$. For example, for $g=h$ the correction factor is 0.84.

\section{Inductive Impedance}

We saw in the previous section that a small pillbox is characterized
by inductive impedance if the frequency is not very large. This is a
common feature of many small perturbations whose size is much smaller
than the pipe radius (e.g., small holes, shallow obstacles on the
wall, etc.) --- for not very large frequencies their impedance is
purely inductive.

The longitudinal wake corresponding to the inductive impedance can be
found by using Eq. (\ref{wake_through_impedance}):\footnote{Since the
integral involved in the calculation of $w_l(s)$ actually diverges at
$\omega \rightarrow \infty$, it should be treated as a generalized
function. It is easier to verify Eq. (\ref{inductive_wake}) by
putting it into Eq. (\ref{impedancefinitions}) and checking that the
resulting impedance is given by Eq. (\ref{inductive_imped}).}
    \begin{equation}\label{inductive_wake}
    w_{l}(s)={\cal L}\delta'(s).
    \end{equation}

Because of the inductive wake, slices of the beam can change their
energy, although the net energy loss for the bunch is zero because
the real part of the impedance vanishes. We can find the energy
change as a function of position within the bunch by using Eq.
(\ref{energy_change}). Integration gives
    \begin{equation}
    \Delta E(s) = -e^{2}N {\cal L}\lambda'(s).
    \end{equation}
For a Gaussian distribution function this reduces to
    \begin{equation}
    \Delta E(s) = {e^{2}N{\cal L} \over \sigma_{z}^{2}} {\xi
    e^{-\xi^{2}/2}\over \sqrt{2 \pi}},
    \end{equation}
where $\xi=s/\sigma_{z}$. For the rms energy spread we find
    \begin{equation}
    \langle \Delta E^2 \rangle^{1/2}=
    {3^{-3/4}(2\pi)^{-1/2}}
    {e^{2}N{\cal L} \over \sigma_{z}^{2}}.
    \end{equation}

\section{Cavity Impedance}

\begin{figure}[ht]
    \begin{center}
    \includegraphics[scale=0.8]{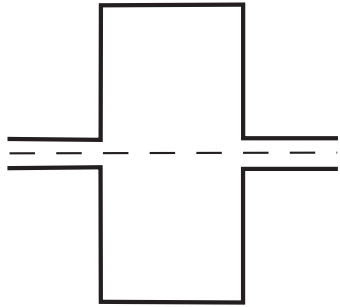}
    \end{center}
    \caption{An RF cavity with beam pipes.\label{cavity}}
    \end{figure}

In the more general case of a large cavity (Fig. \ref{cavity}) the
beam excites cavity eigenmodes and
the longitudinal wake in the cavity is composed of contributions from
single modes,
    \begin{equation}
    w_{l}(s)=\sum_{n} w^{(n)}(s).
    \end{equation}
For perfectly conducting walls, assuming that the modes do not
propagate into the beam pipes,\footnote{This is true if the frequency
of the mode is above the cutoff frequency for the pipe.} the mode
frequencies $\omega_{n}$ are real. It should be no surprise that
each partial wakefield oscillates with the frequency of the mode
$\omega_n$,
    \begin{equation}\label{cavitywake}
    w^{(n)}(s)=2k_{n}\cos\left({\omega_{n}s\over c}\right),
    \end{equation}
where $k_{n}$ is the {\em loss factor}, which depends on the
geometry of the cavity and the mode number. As an example, for a
cylindrical cavity with $b=l$ and $\mathrm{TM}_{010}$ mode,
$k_{010}=4.5/l$. For a more rigorous derivation of the wake for a
cavity, see Ref. 12.

The cavity impedance for this wake can be calculated by using
Eqs. (\ref {impedancefinitions}). They give
    \begin{eqnarray}
    \Re Z_{l} &=&\pi k_{n}[\delta (\omega+\omega_{n})+\delta
    (\omega-\omega_{n})],\nonumber
    \\
    \Im Z_{l} &=&
    k_{n}\left[\frac{1}{\omega+\omega_{n}}+\frac{1}{\omega-\omega_{n}}\right].
    \end{eqnarray}

It is also easy to generalize the above wake for a cavity with
lossy walls when the frequency of the mode has a \emph{small}
imaginary part $\gamma_n$,  $\gamma_n \ll \omega_n$. The wake now decays with time as
    \begin{equation}\label{cavitywake1}
    w_{l}(s)=2k_{n}e^{-\gamma s/c}\cos\left({\omega_{n}s\over c}\right).
    \end{equation}
Again using Eq. (\ref {impedancefinitions}), we can calculate the
impedance. It has two peaks: one in the vicinity of
$\omega=\omega_{n}$ and the other in the vicinity of
$\omega=-\omega_{n}$. Assuming that $\omega$ is close to
$\omega_{n}$, we find
    \begin{equation}
    Z_{l}={ k_{n} \over {\gamma-i
    \left(\omega-\omega_{n}\right)}}.
    \end{equation}



\section*{ APPENDIX}

\renewcommand{\theequation}{\mbox{A\arabic{equation}}}
\setcounter{equation}{0}

\begin{figure}[ht]
    \begin{center}
    \includegraphics[scale=0.8]{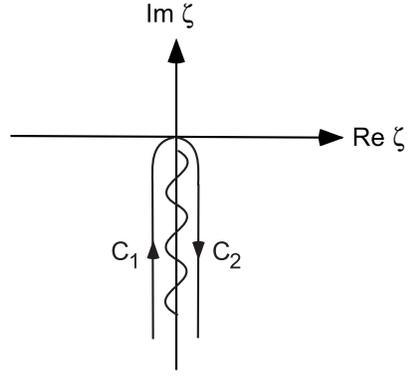}
    \end{center}
    \caption{Complex plane $\zeta$. The wavy line indicates the
    cut. The deformed integration contour consists of two paths, $C_1$ and
    $C_2$, corresponding to the integrals $I_1$ and $I_2$,
    respectively.\label{complex_plane}}
    \end{figure}

We show here how to calculate the integral in Eq. (\ref{eq17}),
    \begin{equation}
    I(\xi)=\int_{-\infty}^{\infty}
    d\omega \sqrt{\omega}e^{-i\omega\xi},
    \end{equation}
where $\xi=t-z/c=s/c$. First, we change the integration variable
$\omega$ to $\zeta=\omega\xi$,
    \begin{equation}
    I(\xi)=\frac{1}{\xi^{3/2}}\int_{-\infty}^{\infty}
    d\zeta \sqrt{\zeta}e^{-i\zeta}.
    \end{equation}
We then consider $\zeta$ as a complex variable. In order to make the
integrand a single-valued function in the complex plane, we make a
cut along the negative imaginary half-axis and deform the integration
path from a straight line to a contour, shown in Fig. \ref{complex_plane}.
The integral $I(\xi)$ can now be split into two parts, $I_1(\xi)$ and
$I_2(\xi)$, corresponding to the left and right branches of the
integration contour.

In the first integral $I_1(\xi)$ we change the complex integration
variable $\zeta$ to the real positive variable $\tau$, $\zeta =
e^{3\pi i/2} \tau = -i\tau$, so that $\sqrt{\zeta} = e^{3\pi i/4}
\sqrt{\tau} = (i-1)\sqrt{\tau}/\sqrt{2}$. This gives for $I_1$
    \bea
    I_1(\xi)&=&\frac{1}{\xi^{3/2}}\int_{\infty}^{0}
    (-i d\tau) \frac{(i-1)\sqrt{\tau}}{\sqrt{2}}e^{-\tau}
    \nonumber\\
    &=&\frac{(i+1)}{\sqrt{2}\xi^{3/2}}\int_{\infty}^{0}
    d\tau \sqrt{\tau}e^{-\tau}
    \nonumber\\
    &=&-\frac{\sqrt{\pi}(i+1)}{2^{3/2}\xi^{3/2}}.
    \eea
For the second integral $I_2(\xi)$ we choose the real positive
variable $\tau$ such that $\zeta = e^{-\pi i/2} \tau = -i\tau$, which
means that $\sqrt{\zeta} = e^{-\pi i/4} \sqrt{\tau} =
(1-i)\sqrt{\tau}/\sqrt{2}$ and
    \bea
    I_2(\xi)&=&\frac{1}{\xi^{3/2}}\int_{0}^{\infty}
    (-i d\tau) \frac{(1-i)\sqrt{\tau}}{\sqrt{2}}e^{-\tau}
    \nonumber\\
    &=&I_1(\xi).
    \eea
Hence
    \begin{equation}
    I(\xi)=I_1(\xi)+I_2(\xi)
    =-\frac{\sqrt{\pi}(i+1)}{2^{1/2}\xi^{3/2}}.
    \end{equation}
Substituting this equation into Eq. (\ref{eq17}) gives Eq.
(\ref{Ez_lossy_wall}).

\end{document}